\documentclass[journal]{IEEEtran}
\pdfoutput=1

\usepackage{amssymb}
\usepackage{amsfonts}
\usepackage{bbm}
\usepackage{amsfonts}
\usepackage[dvips]{graphicx}
\usepackage[dvips]{color}
\usepackage{graphicx}
\usepackage{amsmath}
\usepackage{fancyhdr}
\usepackage{stfloats}
\usepackage{multirow}
\usepackage{algorithm}
\usepackage{algorithmic}
\usepackage{cite}
\usepackage[bookmarks=false]{}
\usepackage{amsthm}
\usepackage{algorithm}
\usepackage{algorithmic}
\usepackage{mathbbol}
\usepackage{amsfonts}
\usepackage{graphicx}
\usepackage{amsmath}
\usepackage{amssymb}
\usepackage{latexsym}
\usepackage{graphicx}
\usepackage{cite}
\usepackage{url}
\usepackage{stfloats}
\usepackage{mathrsfs}
\usepackage{setspace}
\usepackage{booktabs}
\usepackage{latexsym}
\usepackage{url}
\usepackage{stfloats}
\usepackage{mathrsfs}
\theoremstyle{plain}

\usepackage{cite}
\usepackage{setspace}
\usepackage{textcomp}
\usepackage{xcolor}
\usepackage{makecell}
\usepackage{etoolbox}
\usepackage{diagbox}
\usepackage{caption}
\usepackage{amsthm}
\usepackage{multirow}
\usepackage{array}
\usepackage{colortbl}
\usepackage{bbding}
\usepackage[tight,footnotesize]{subfigure}
\usepackage[]{hyperref}
\hypersetup{
	bookmarksnumbered=true,     
	bookmarksopen=true,         
	bookmarksopenlevel=1,       
	colorlinks=true,
	linkcolor=blue            
}
\begin{document}
\clearpage
\title{\huge Explainable Semantic Communication for Text Tasks}
%

\author{Chuanhong Liu, Caili Guo, \emph{Senior Member}, \emph{IEEE}, Yang Yang, Wanli Ni, Yanquan Zhou, Lei Li, \\and Tony Q.S. Quek, \emph{Fellow}, \emph{IEEE}
\thanks{This work was supported in part by the Fundamental Research Funds for the Central Universities (No.2021XD-A01-1), and in part by BUPT Excellent Ph.D. Students Foundation (No.CX2022101). An earlier version of this paper \cite{Liu_Triplet} was presented in part at the 2023 IEEE Wireless Communications and Networking Conference (WCNC) [DOI: 10.1109/WCNC55385.2023.10118916]. (\textit{Corresponding author: Caili Guo})}
\thanks{Chuanhong Liu and Caili Guo are with the Beijing Key Laboratory of Network System Architecture and Convergence, School of Information and Communication Engineering, Beijing University of Posts and Telecommunications, Beijing 100876, China (e-mail: 2016\_liuchuanhong@bupt.edu.cn; guocaili@bupt.edu.cn).}
\thanks{Yang Yang is with the Beijing Laboratory of Advanced Information Networks, School of Information and Communication Engineering, Beijing University of Posts and Telecommunications, Beijing 100876, China (e-mail: yangyang01@bupt.edu.cn).}
\thanks{Wanli Ni is with Department of Electronic Engineering, Tsinghua University, Beijing 100084, China (e-mail: niwanli@tsinghua.edu.cn).}
\thanks{Yanquan Zhou and Lei Li are with School of Artificial Intelligence, Beijing University of Posts and Telecommunications, Beijing 100876, China (e-mail: zhouyanquan@bupt.edu.cn; leili@bupt.edu.cn).}
\thanks{Tony Q. S. Quek is with the Dept. of Information Systems Technology and Design, Singapore University of Technology and Design, Singapore, 487372 (e-mail: tonyquek@sutd.edu.sg).}
}

\maketitle
\pagestyle{headings}
\vspace{0cm}
\begin{abstract}
Task-oriented semantic communication has gained increasing attention due to its ability to reduce the amount of transmitted data without sacrificing task performance. Although some prior efforts have been dedicated to developing semantic communications, the semantics in these works remains to be unexplainable. Challenges related to explainable semantic representation and knowledge-based semantic compression have yet to be explored. In this paper, we propose a triplet-based explainable semantic communication (TESC) scheme for representing text semantics efficiently. Specifically, we develop a semantic extraction method to convert text into triplets while using syntactic dependency analysis to enhance semantic completeness. Then, we design a semantic filtering method to further compress the duplicate and task-irrelevant triplets based on prior knowledge. The filtered triplets are encoded and transmitted to the receiver for completing intelligent tasks. Furthermore, we apply the propsed TESC scheme to two emblematic text tasks: sentiment analysis and question answering, in which the semantic codec is meticulously customized for each task. Experimental results demonstrate that 1) TESC scheme outperforms benchmarks in terms of Top-1 accuracy and transmission efficiency, and 2) TESC scheme enjoys about 150\% performance gain compared to the traditional communication method.
\end{abstract}


\section{Introduction}
\label{sec:intro}

\IEEEPARstart{T}{he} {advancement of 6G network technology is increasingly intertwined with artificial intelligence (AI)\cite{Walid_6G, Chen_distribute, Letaief_Roadmap}. As the number of connected intelligent devices skyrockets and wireless data traffic grows exponentially, one of the primary challenges emerging is spectrum scarcity. Traditional communication technologies typically focus on the accurate transmission of symbols but often overlook the actual purpose and meaning of the data being transmitted. This leads to inefficient use of wireless communication resources and struggles to meet the demands of future large-scale communications. Semantic communication, recognized as a key technology in 6G networks, offers a promising solution\cite{Qin_survey, Nine}.}

{Unlike traditional communication that transmits raw data, semantic communication emphasizes transmitting task-oriented semantics\cite{framework_Yang, info}. This paradigm shift has multiple advantages. First, it reduces redundant or irrelevant content, significantly improving communication efficiency and alleviating the spectrum scarcity issue. Second, it enhances the robustness of communication systems against channel noise by allowing bit-level errors. However, the unexplainable nature of existing semantic communication hinders its practical application to some extent. Therefore, it is necessary to improve its explainability so as to enhance the reliability of semantic communication systems.}

\vspace{-0.2cm}
\subsection{Related Works}
{Recently, there are a number of prior studies on semantic communications \cite{Xie_Deep, Xie_lite, Gunduz_JSCC, Weng1}. In particular, for text, the authors in \cite{Xie_Deep} proposed a semantic communication system based on Transformer, in which the concept of semantic information was clarified at the sentence level. Based on \cite{Xie_Deep}, the authors in \cite{Xie_lite} further proposed a lite distributed semantic communication system, making the model easier to deploy on the Internet of Things (IoT) devices. For image, the authors in \cite{Gunduz_JSCC} presented a joint source-channel coding scheme based on convolutional neural networks (CNN) to transmit image data in a wireless channel, which can jointly optimize various modules of the communication system. For speech, the authors in \cite{Weng1} proposed an attention mechanism-based semantic communication system for speech signals, in which a general model to cope with various channel conditions without retraining is developed. All of them concentrate solely on reconstructing source data at the receiver, neglecting the specific task at hand. This oversight may significantly restrict the data compression efficiency.}

\begin{figure}[t]
\centering
\subfigure[Feature-based unexplainable semantic representation]{
\includegraphics[width=0.9\linewidth]{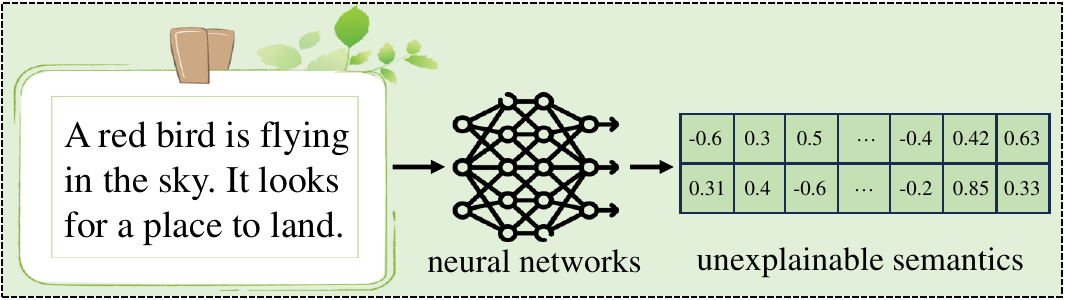}}
\subfigure[Triplet-based explainable semantic representation]{
\includegraphics[width=0.9\linewidth]{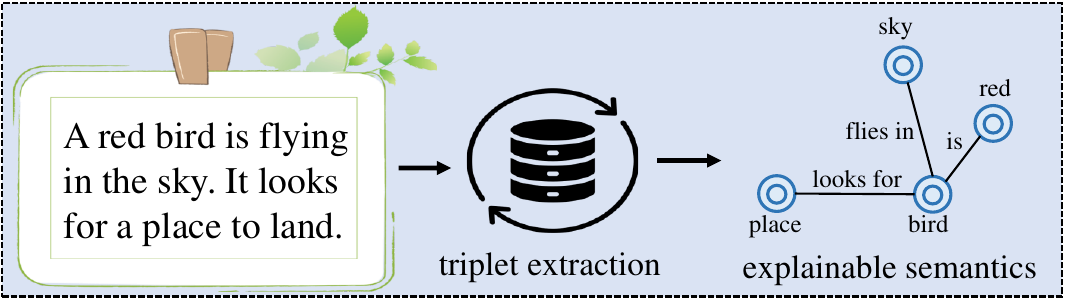}}
\caption{Unexplainable vs. explainable semantic representation.}
\label{fig:sem_pre}
\vspace{-0.3cm}
\end{figure}

{Focusing on specific mission, task-oriented semantic communication has been proposed in \cite{retrival, Lee, liu2023adaptable, wei2023federated, MU-DeepSC1}. The authors in \cite{retrival} considered image-based re-identification for persons or cars as the communication task, where two schemes were proposed to improve the retrieval accuracy. The authors in \cite{Lee} designed a joint transmission-classification system for images, in which the receiver outputs image classification results directly. The authors in \cite{liu2023adaptable} proposed an adaptable semantic compression method for image classification-oriented semantic communication system, and investigated resource allocation for performance optimization. The authors in \cite{wei2023federated} presented a novel federated semantic learning framework to collaboratively train the semantic-channel encoders of multiple devices with the coordination of a base station-based semantic-channel decoder, aiming at semantic knowledge graph constructing task. The authors in \cite{MU-DeepSC1} proposed a multi-users semantic communication system for serving visual question answering task, in which long short term memory was used for text transmitter and CNN for image transmitter. All of the aforementioned studies employed deep neural networks to encode data and utilized the output features as the semantic representation of the data. However, the output of neural networks, which consists of a vector of numerical values, is often unexplainable and lacks the logic of human language, as shown in Fig. \ref{fig:sem_pre}(a).}

{In \cite{Uysal, Kountouris, Ma_explainable, Jiang_KG, Hu_robust, Wang_Attention2}, the authors proposed a number of methods for explainable semantic communications. The authors in\cite{Uysal} and \cite{Kountouris} proposed a semantic communication framework tailored for real-time control systems, where the control signals were treated as the semantic information of the data. However, it should be noted that utilizing control signals as semantic information may not be suitable for transmitting text or image data, as control signals are unable to capture the content associated with textual or visual information. The authors in \cite{Ma_explainable} proposed an explainable and robust semantic communication framework to disentangles features into independent and semantically interpretable features. However, this work is specifically dedicated to the field of image transmission and does not encompass text-related applications. Moreover, the authors in \cite{Jiang_KG, Hu_robust, Wang_Attention2} proposed to use the knowledge graph instead of features to represent text semantics, which could decompose texts into multiple triplets. The entities in such a representation method were well-defined without ambiguity, since knowledge graphs were usually used as generic databases, and the information contained in them was clear and general. However, the textual statements in daily language depend on each other, and most of the words are only applicable to a specific context. A large number of ambiguous words could not be extracted completely via the existing methods in \cite{Jiang_KG, Hu_robust, Wang_Attention2}. In addition, all of these works aimed at recovering the original text at the receiver, rather than implementing a specific task, which may lead to some task-irrelevant semantic redundancy.}

\subsection{Challenges and Contributions}
The development of task-oriented explainable semantic communication systems faces a number of challenges. The first one is to represent the task-oriented semantics of text in an interpretable way, as shown in Fig. \ref{fig:sem_pre}(b). This involves not only recognizing the meaning of the text, but also ensuring that this representation is understandable. Furthermore, it is crucial to design an efficient method to extract these interpretable semantics. This process must be robust and efficient, enabling the system to extract the semantics accurately. Another challenge involves refining the extracted semantics, especially in terms of removing semantic redundancy. It is crucial for communication systems to transmit only the critical semantics as this helps to improve the overall efficiency of the system.

In this paper, we propose a novel triplet-based explainable semantic communication (TESC) scheme for text tasks, which can effectively filter the redundant information to reduce the amount of transmitted data. The main contributions of this paper are summarized as follows:

\begin{itemize}
    \item[$\bullet$] We propose a TESC scheme for text tasks, where the transmitter characterizes the semantic information of the text by means of triplets, which enhances the interpretability of the semantic information while efficiently completing the downstream tasks. Specifically, the triplet-based semantic representation method includes semantic extraction and semantic filtering. The former consists of an out-of-the-box triplet extraction tool and a complementary triplet extraction method that employs syntactic dependency analysis. This dual approach ensures a comprehensive capture of semantic information. The latter is a two-step method for compressing the extracted triplets, thereby reducing the amount of transmitted data and improving the efficiency of semantic communication.
    \item[$\bullet$] We apply the proposed TESC scheme to two emblematic text tasks: sentiment analysis and question answering (QA), underscoring its effectiveness and versatility. The semantic codec is meticulously customized for each task, optimizing for nuanced understanding and processing, thereby exemplifying the scheme's potential across diverse text tasks.
    \item[$\bullet$] Experiment results show that the proposed TESC scheme can significantly reduce the amount of transmitted data and improve task performance, compared to the traditional communication and existing semantic communication methods. Specifically, we observe that 1) the proposed TESC scheme shows 80.5\% and 150\% performance gain over traditional communication methods on sentiment analysis and QA, respectively. 2) The number of transmission symbols of TESC is only 8\% of that of traditional methods in QA taks. 3) The proposed semantic filtering method notably decreases the average number of triplets by as much as 76.1\% and the total word count by 83.8\% while maintaining task performance.
\end{itemize}

The remainder of this paper is organized as follows. The proposed TESC scheme is described in Section \ref{sec:system}. The proposed TESC scheme is applied to sentiment analysis and QA task in Section \ref{sec:Application}. Finally, experiments results are provided in Section \ref{sec:simulation}, which is followed by conclusions in Section \ref{conclusion}.



\section{Proposed TESC Scheme}
\label{sec:system}

\subsection{Scheme Design}
\begin{figure*}[t]
	\begin{center}
		\includegraphics[width=1\linewidth]{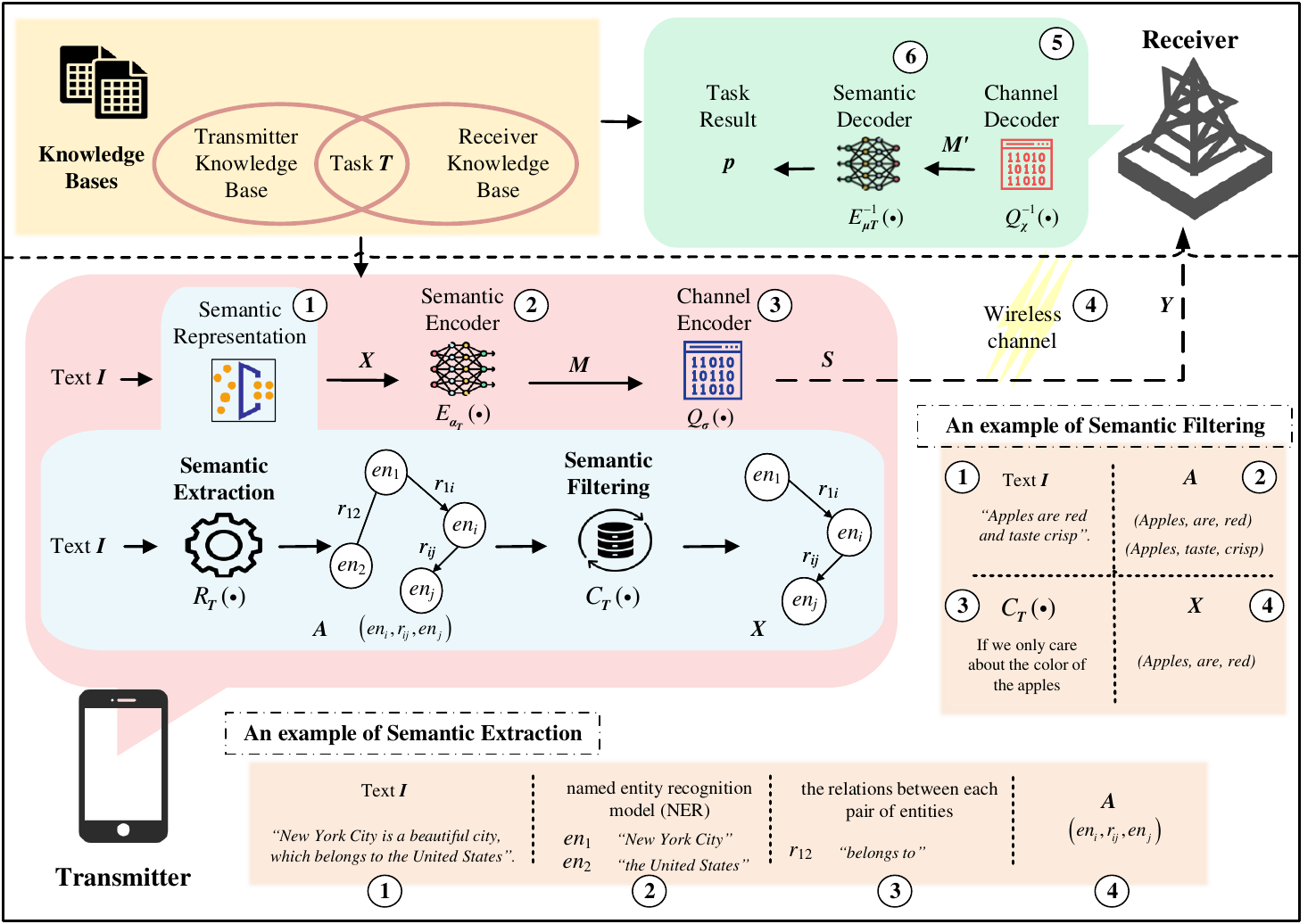}
	\end{center}
	\caption{An illustration of the proposed TESC scheme for text tasks.}
	\label{fig:model}
\end{figure*}

As shown in Fig. \ref{fig:model}, the structure of the proposed TESC scheme consists of a semantic representation module, a semantic codec module, a channel codec module and knowledge bases. {Similar to \cite{niu2022paradigm}, the knowledge base primarily contains well-trained model parameters, alongside prior task-specific knowledge.} In this paper, the channel codec is the same as that of the conventional communication system, we mainly focus on the design of semantic representation method and semantic codec.

The transmitter aims at gathering data locally and performing an inference task with the assistance of the receiver, which is the goal of semantic communication. In particular, the transmitter first extracts the triplets from original text and further filters the redundant triplets. Then, the triplets are encoded and transmitted to the receiver in a scheduled manner. Finally, the receiver performs semantic decoding based on the received semantics and returns the result of tasks to the transmitter. In the following, we detail the system design of the proposed TESC scheme.

As shown in Fig. \ref{fig:model}, the transmitter maps the input text into complex symbol streams, and then passes it through the wireless channel with fading and noise. Particularly, the input of the system is denoted by ${\boldsymbol I}{\rm{ = }}\left[ {{i_1},{i_2}, \cdots ,{i_N}} \right]$, where $i_n$ represents the $n$-th word in the text and $N$ is the number of words in ${\boldsymbol I}$. The transmitter first extracts triplets from original text ${\boldsymbol I}$ to represent its semantics, which can be denoted by
\begin{eqnarray}\label{signalform}
	{\boldsymbol {A}} = {R_{\boldsymbol {T}}}({\boldsymbol {I}}),
\end{eqnarray}
where ${R_{\boldsymbol {T}}}(\cdot)$ denotes the semantic extraction method, and ${\boldsymbol {T}}$ is the oriented task. The output ${\boldsymbol {A}}$ is a series of triplets, consisting of entities and relations.

In particular, each entity in the triplets refers to an object or a concept in the real world. Hereinafter, we define entity $i$ in text ${\boldsymbol I}$ as $en_i$ that consists of a subset of words in text ${\boldsymbol I}$. For example, ``\emph{New York City}'' and ``\emph{the United States}" are two entities consisting of three words in the text ``\emph{New York City is a beautiful city, which belongs to the United States}". There are various methods such as named entity recognition model (NER) can be used to extract the entities in text ${\boldsymbol I}$. Given a pair of extracted entities ($en_i, en_j$), the semantic extractor has to extracted the relation $r_{ij}$ between them. For example, the relation between entity ``\emph{New York City}" and entity ``\emph{the United States}" can be denoted by ``\emph{belongs to}". Based on the extracted entities and relations, the triplets of text ${\boldsymbol I}$ can be expressed as
\begin{eqnarray}\label{}
{\boldsymbol{A}}{\rm{ = [}}{{\boldsymbol{a}}_{\rm{1}}}{\rm{,}}...{\rm{,}}{{\boldsymbol{a}}_{{k}}}{\rm{,}}...{\rm{,}}{{\boldsymbol{a}}_{{K}}}{\rm{]}},
\end{eqnarray}
where ${{\boldsymbol{a}}_k} = (e{n_{k,i}},{r_{k,ij}},e{n_{k,j}})$ is the $k$-th triplet and $K$ is the total number of triplets extracted from ${\boldsymbol I}$. Note that the relations are directional, and hence we have ${r_{k,ij}} \neq {r_{k,ji}}$. The triplet-based representation can effectively remove a significant amount of semantic redundancy from the original text, and thus the data size of the extracted triplets is smaller than that of the original text ${\boldsymbol I}$. The reduction of data size is advantageous for applications that require efficient transmission and understanding of large amounts of text-based data. {It is worth noting that the semantics of original texts is extracted in an explainable manner. This represents a fundamental distinction between our work and existing feature-based semantic representations.}

Different from the conventional communication system, in which bits or symbols are treated equally, triplets may have different semantic importance for accomplishing tasks, and there may still be some redundancy in the extracted triplets\cite{openie1}. Therefore, we can further filter the extracted triplets according to their contribution to the downstream task. 
By retaining only the essential triplets under the guidance of task ${\boldsymbol T}$, the task-irrelevant information is eliminated, which can be especially valuable in scenarios where data transmission is limited by bandwidth or other constraints. The semantic filtering process can be expressed as
\begin{eqnarray}\label{}
	{\boldsymbol{X}} = {C_{\boldsymbol{T}}}({\boldsymbol{A}}),
\end{eqnarray}
where ${C_{\boldsymbol{T}}}(\cdot)$ denotes the semantic filtering function, which highly depends on the final task $\boldsymbol{T}$. {The term $\boldsymbol{X}$ is the task-oriented triplets after filtering, which is a subset of the original triplets $\boldsymbol{A}$, i.e., ${\boldsymbol{X}} \subset {\boldsymbol{A}}$.} Obviously, semantic filtering offers two major benefits by removing semantic redundancy. First, it reduces the amount of transmitted data and consequently reduces transmission delay and the demand for bandwidth. Second, it lessens the transmission energy consumption, which is significant for lightweight devices.

The filtered triplets are encoded via a semantic encoder, which can further compress the transmitted data. The implementation details of the semantic encoder are closely related to the downstream tasks ${\boldsymbol T}$. The encoded semantics can be denoted by
\begin{eqnarray}\label{}
	{\boldsymbol{M}} = {E_{\boldsymbol{\alpha_T}}}({\boldsymbol{X}}),
\end{eqnarray}
where ${E_{\boldsymbol{\alpha_T}}}(\cdot)$ denotes the semantic encoder with parameter set ${\boldsymbol{\alpha_T}}$ for accomplishing task ${\boldsymbol T}$. {The above semantic representation and semantic encoder serve a purpose akin to traditional source coding, which involves compressing the source to enhance wireless communication efficiency.}

Next, the features are encoded by the channel encoder to generate symbols for transmission, which can be denoted by
\begin{eqnarray}\label{}
	{\boldsymbol{S}} = {Q_{\boldsymbol{\sigma}}}({\boldsymbol{M}}),
\end{eqnarray}
where $Q_{\boldsymbol{\sigma}}(\cdot)$ denotes the channel encoder network with parameter set ${\boldsymbol{\sigma}}$.

Then, the encoded symbols are transmitted via a wireless channel, and the received signal is expressed as
\begin{eqnarray}\label{}
{\boldsymbol{Y}} = h{\boldsymbol{S}} + \boldsymbol{n},
\end{eqnarray}
where $h$ is the channel gain and $\boldsymbol{n}$ is the additive white Gaussian noise (AWGN) sampled from ${{\cal C}{\cal N}}(0,{\boldsymbol{\sigma }^2\boldsymbol{I}})$. {For end-to-end (E2E) training of the semantic transceiver, the channel must allow the backpropagation of parameters, and thus physical channel in this work is modeled by a non-trainable fully connected layer in line with \cite{Xie_Deep, Xie_lite, Gunduz_JSCC} to bolster the robustness.}

As shown in Fig. \ref{fig:model}, the receiver includes a channel decoder and a semantic decoder to recover the transmitted semantics and complete the final task. The received symbols are decoded to recover semantics via the channel decoder, which can be expressed as
\begin{eqnarray}\label{}
	{{{\boldsymbol{M}}'}} = {Q_{\boldsymbol{\chi }}^{ - 1}}({\boldsymbol{Y}}),
\end{eqnarray}
where ${Q_{\boldsymbol{\chi }}^{ - 1}}(\cdot)$ denotes the channel decoder with parameter set ${\boldsymbol{\chi }}$.

Finally, the receiver inputs the recovered semantics ${{\boldsymbol{M}}'}$ into the semantic decoder to complete the intelligent tasks. Specifically, the output is
\begin{eqnarray}\label{}
	{\boldsymbol{p}} = {E_{\boldsymbol{\mu_T}}^{ - 1}}({\boldsymbol{{{\boldsymbol{M}}'}}}),
\end{eqnarray}
{where ${\boldsymbol{p}}$ is the task result, which will be returned to the transmitter} and ${E_{\boldsymbol{\mu_T}}^{ - 1}}(\cdot)$ denotes the semantic decoder with the parameter set ${\boldsymbol{\mu_T}}$. {Notably, the entire semantic communication network can be trained offline and subsequently deployed online, resulting in substantial resource savings for devices.}

As stated in \cite{Shi}, accurate and efficient recognition and extraction of semantic information, including identifying different types of entities and their potential relationships, is crucial for effective semantic communication. However, the effective representation of semantic information in TESC poses two primary challenges. One is effectively extracting the triplets from the input text. This requires techniques that can identify and disambiguate the semantic relationships between words and phrases in the text, as well as handle the inherent ambiguity in natural language. Another challenge is filtering the extracted triplets to ensure that they are relevant and informative for the downstream task. Different downstream tasks may require different levels of granularity and specificity in the extracted triplets, and it can be challenging to identify the most useful and relevant triplets for a given task. This requires techniques that can filter out irrelevant or noisy triplets based on the downstream task. In the next subsection, we will present the proposed semantic representation method in detail.

\begin{figure*}[t]
	\begin{center}
		\includegraphics[width=0.9\linewidth]{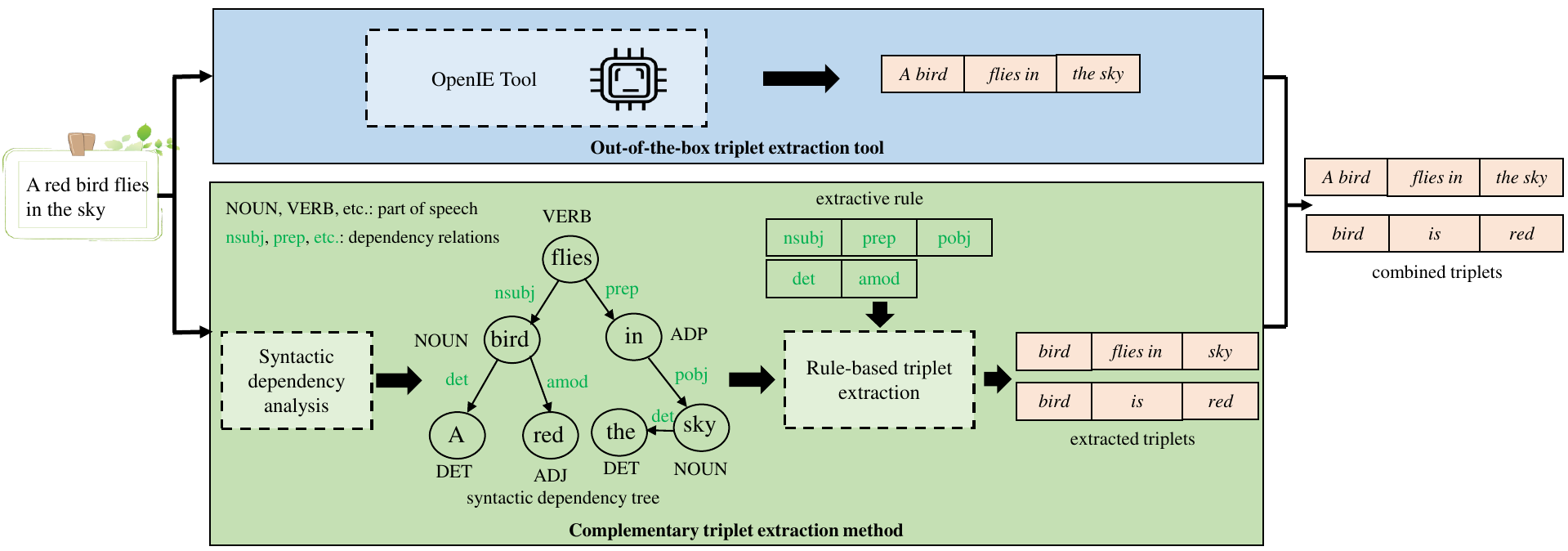}
	\end{center}
	\caption{The process of the proposed semantic extraction method.}
	\label{fig:extraction}
\end{figure*}

\subsection{Triplet-Based Semantic Representation Method}
In this subsection, we delve into the proposed triplet-based semantic representation method, which comprises of two key components: semantic extraction and semantic filtering. First, we introduce syntactic dependency analysis-based semantic extraction method. Then, we elaborate the details of the semantic filtering method under the task guidance.

\subsubsection{Semantic Extraction Method}
{The proposed semantic extraction method, depicted in Fig. \ref{fig:extraction}, leverages both an established out-of-the-box triplet extraction tool and a novel complementary triplet extraction method predicated on syntactic dependency analysis. In particular, the out-of-the-box triplet extraction tool used in this work is the Open Information Extraction (OpenIE) annotator\cite{openie2}. {However, the existing triplets extraction methods still exhibit moderate performance, and this limitation could potentially be the performance bottleneck in semantic communication. To overcome this problem, we further propose a complementary triplet extraction method based on syntactic dependency analysis to supplement the missing semantics. Syntactic dependency analysis offers a comprehensive depiction of the dependency relationship between linguistic elements at the sentence level, making it particularly advantageous for extracting implicit and abstract information.} For example, OpenIE can only extract (\textit{a bird}, \textit{flies in}, \textit{sky}) from the text ``\textit{A red bird flies in sky}'', while the proposed complementary triplet extraction method can additionally extract more specific triplets (\textit{bird}, \textit{is}, \textit{red}).}

{The complementary triplet extraction method can be divided into two phases: syntactic dependency analysis ($\boldsymbol{S}_{\rm{d}}$) and rule-based triplet extraction ($\boldsymbol{E}_{\rm{r}}$), as shown in the lower part of Fig. \ref{fig:extraction}. In the first phase, we employ Spacy to deconstruct the sentence syntactically, outputting a tree whose branches represent the syntactic dependencies. Spacy is a powerful open-source natural language processing (NLP) library renowned for its efficiency and accuracy in linguistic analysis. Mathematically, this can be represented as}
{
\begin{eqnarray}\label{}
	{\boldsymbol{A}_{\rm{d}}} = \boldsymbol{S}_{\rm{d}}({\boldsymbol{I}}),
\end{eqnarray}
{where $\boldsymbol{I}$ denotes the input texts, and ${\boldsymbol{A}_{\rm{d}}}$ represents the dependency tree. The nodes of ${\boldsymbol{A}_{\rm{d}}}$ represent the words in the sentence and the edges represent the syntactic relationships between the words.}}

{In rule-based triplet extraction phase, we apply a set of predefined rules to ${\boldsymbol{A}_{\rm{d}}}$ to extract triplets ${\boldsymbol{A}_{\rm{c}}}$, which can be denoted by
\begin{eqnarray}\label{}
	{\boldsymbol{A}_{\rm{c}}} = \boldsymbol{E}_{\rm{r}}({\boldsymbol{A}_{\rm{d}}}).
\end{eqnarray}}
where ${\boldsymbol{A}_{\rm{c}}}$ is the set of complementary triplets.

{These rules are designed to capture various sentence structures, including, but not limited to, the classic subject-verb-object and subject-link verb-predicative structures. For instance, in the case of a subject-verb-object structure, we identify all the verbs in the syntactic dependency tree and assign them as the relationship in the extracted triplets. The subject node and object node connected to a specific verb are then labeled as the head entity and tail entity of the triplet, respectively. Additionally, we expand the sub-tree of the extracted node and include all its modifiers, which can be abstract and non-generic, in the corresponding triplets to ensure semantic completeness.}


{Finally, the extracted triplets of the two parts will be combined and the duplicate triplets will be sifted out. Based on the proposed method, the semantics of the raw text, including the abstract and non-generic information, can be fully represented.}

\subsubsection{Semantic Filtering Method}
{During the triplet extraction process, redundant triplets that are either semantically identical or irrelevant to the task may be generated, presenting a challenge to communication efficiency. This redundancy can lead to an extensive quantity of extracted triplets. {Leveraging the explainability of triplets, we propose a two-step explicit semantic filtering method to address this issue.} Table \ref{tab:filtering} presents an example of how the proposed two-step semantic filtering method works.}

{In the first step, we apply a unique-relationship filter, $\boldsymbol{F}_{\rm{u}}$, to remove semantic-redundant triplets. This approach is grounded in the principle that the relationship between two entities is unique\cite{Che_KG}. Specifically, we begin with an empty set initialized for entity pairs. As we extract each triplet, we examine whether the pair of its head entity and tail entity is already included in the set. If this specific entity pair is found in the set, we discard the corresponding triplet. Conversely, if the pair is novel, we incorporate it into our list of triplets. This preliminary processing step helps to significantly reduce the duplicate semantics transmitted, which can be denoted by
\begin{eqnarray}\label{}
	{\boldsymbol{X}_{\rm{u}}} = \boldsymbol{F}_{\rm{u}}({\boldsymbol{A}}),
\end{eqnarray}
where ${\boldsymbol{X}_{\rm{u}}}$ is the remain triplets after filtering, and ${\boldsymbol{A}}$ is the set of original extracted triplets.} As depicted in Table \ref{tab:filtering}, once the triplet (``\textit{China}", ``\textit{capital city}", ``\textit{Beijing}") is extracted, the entity pair (``\textit{China}",``\textit{Beijing}") is added to the entity pair set. Hence, when the triplet (``\textit{China}", ``\textit{contain}", ``\textit{Beijing}") appears, it will be discarded, as the entity pair (``\textit{China}",``\textit{Beijing}") already exists in the set.

In the second step, we employ a task-specific filter, $\boldsymbol{F}_{\rm{t}}$, informed by the knowledge base, to discard triplets irrelevant to the task, which can be represented as
\begin{eqnarray}\label{}
	{\boldsymbol{X}_{\rm{t}}} = \boldsymbol{F}_{\rm{t}}({\boldsymbol{X}_{\rm{u}}}),
\end{eqnarray}
where ${\boldsymbol{X}_{\rm{t}}}$ is the final transmitted semantic triplets. The knowledge-based semantic filtering step is closely tied to the task at hand, and we can illustrate this through examples such as sentiment analysis and QA tasks. In sentiment analysis, we can leverage the knowledge that triplets with longer entities and richer relationships tend to include more verbs or adjectives expressing sentiment. These triplets are more valuable for analyzing the sentiment of the original text. Therefore, after filtering redundant triplets in the first step, there is a preference for retaining triplets with longer entities and richer relationships. In addition, for QA task, we possess prior knowledge that questions and answers typically involve specific types of entities, such as position, size, time, etc. Thus, unmatched or irrelevant triplets have a higher likelihood of being filtered out. Mathematically, we employ a selection score ${S_{\rm{QA}}}\left( {{{\boldsymbol{a}}_i}} \right)$ calculated as follows
\begin{eqnarray}\label{}
{S_{{\rm{QA}}}}\left( {{{\boldsymbol{a}}_i}} \right) = \sum\limits_{t \in \mathcal{T}} {\delta \left( {{{\boldsymbol{a}}_i},t} \right)},
\end{eqnarray}
where $\mathcal{T}$ is a entity set of specific type. $\delta \left( \cdot \right)$ is an indicator function that returns 1 if the triplet ${\boldsymbol{a}}_i$ contains an entity belonging to $\mathcal{T}$ and 0 otherwise. Triplets with a selection score of 0 will be filtered out, while others will be retained. This scoring mechanism ensures that triplets directly pertinent to the question's context are prioritized for retention. As depicted in step 2 of Table \ref{tab:filtering}, when the question is about country, the selection score of triplet (``\textit{Bob}", ``\textit{born in}", ``\textit{Beijing}") is 0 and it will be removed.

Semantic filtering method allows for a nuanced and precise refinement of data, tailored to the specific requirements of intelligent tasks, thereby ensuring that TESC scheme not only minimizes data transmission but does so without compromising the integrity and applicability of the retained information. Moreover, some experiments have been conducted in Section \ref{sec:simulation}, the results of which substantiate the effectiveness of the proposed semantic filtering method.


\begin{table}[t]
	\normalsize
	\centering
	\caption{Example of Semantic Extraction and Semantic Filtering.}
	\setlength{\abovecaptionskip}{-0.5cm}
        \renewcommand\arraystretch{1.5}
        \resizebox{\linewidth}{!}{
        \begin{tabular}{|p{3cm}<{\centering}|p{3cm}<{\centering}|p{3cm}<{\centering}|}
            \hline
            \rowcolor{red!10} \multicolumn{3}{|c|}{\textbf{Original Texts}}\\
		\hline
		 \multicolumn{3}{|c|}{Beijing is the capital city of China, and Bob was born in there.}\\
            \hline
		\rowcolor{red!10} \multicolumn{3}{|c|}{\textbf{Extracted Results}}\\
            \hline
            \rowcolor{red!10} \textbf{Subject} & \textbf{Relation} & \textbf{Object}\\
            \hline
             China & capital city & Beijing\\
            \hline
             China & contain & Beijing\\
            \hline
             Bob & born in & Beijing\\
            \hline
		\rowcolor{red!10} \multicolumn{3}{|c|}{\textbf{Filtered Results (Step 1)}}\\
            \hline
            \rowcolor{red!10} \textbf{Subject} & \textbf{Relation} &\textbf{Object}\\
            \hline
             China & capital city & Beijing\\
            \hline
             Bob & born in & Beijing\\
            \hline
        \rowcolor{red!10} \multicolumn{3}{|c|}{\textbf{Filtered Results (Step 2)}}\\
            \hline
            \rowcolor{red!10} \textbf{Subject} & \textbf{Relation} &\textbf{Object}\\
            \hline
             China & capital city & Beijing\\
            \hline
	\end{tabular}}
	\vspace{-0.cm}
	\label{tab:filtering}
\end{table}

\vspace{-0.2cm}
\section{Application of TESC Scheme in Sentiment Analysis and Question Answering}
\label{sec:Application}
In this section, we offer a detailed exploration of how our proposed TESC scheme is applied across various text tasks, with a particular focus on sentiment analysis\cite{sentiment} and QA\cite{QA}. These tasks are chosen for their widespread importance in NLP and their unique attributes, which can effectively showcase the superiority of the proposed scheme\cite{nlp}. {Considering in the smart home scenario, sentiment analysis can enhance the interaction between users and smart home devices by interpreting emotional states, while question answering systems can facilitate seamless dialogue with various IoT devices, thus enriching user experience.} TESC scheme can be easily expanded to other intelligent tasks via redesigning the semantic codec. {Additionally, all models can be trained in the cloud and then broadcast to users, making the implementation process more efficient and scalable.}

\subsection{Sentiment Analysis}
\label{sec:SA}
In this subsection, we will focus on sentiment analysis-oriented TESC. Sentiment analysis is a technique used to analyze subjective texts with various emotions, such as positive or negative, to determine the text's opinions, preferences, and emotional tendencies. Sentiment analysis can be divided into binary classification and multi-classification tasks, in which the latter providing a finer granularity in the division of emotions. For the scope of this paper, we will focus on binary sentiment classification. However, it is worth noting that the proposed TESC scheme can be easily applied to multi-classification tasks as well. 


\begin{figure*}[t]
	\begin{center}
		\includegraphics[width=1\linewidth]{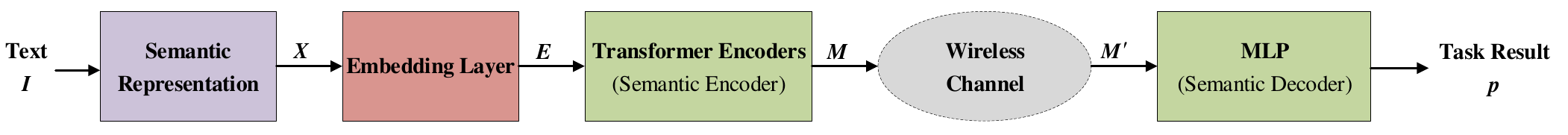}
	\end{center}
	\caption{The process of sentiment analysis-oriented TESC.}
	\label{fig:sentiment}
\end{figure*}

The proposed sentiment analysis-oriented TESC is shown in Fig. \ref{fig:sentiment}. It should be noted that since channel coding and decoding are not the primary focus of this work, they are omitted in the figure. First, we extract the semantics of the raw text ${\boldsymbol I}$ to obtain the triplets ${\boldsymbol X}$, based on the proposed semantic extraction and filtering methods in Sec. III. The triplets are separated by commas and input to the encoder network.

Then, we encode the triplets ${\boldsymbol X}$ to convert them into the embedding ${\boldsymbol E}$, which can be represented as
\begin{eqnarray}\label{}
	{\boldsymbol{E}} = {\boldsymbol{E}}_{T} + {\boldsymbol{E}}_{P} + {\boldsymbol{E}}_{S},
\end{eqnarray}
where ${\boldsymbol{E}}_{T}$ is the token embedding, ${\boldsymbol{E}}_{P}$ represents positional embedding, and ${\boldsymbol{E}}_{S}$ denotes the segmentation embedding. The token embedding represents the index of the token in the vocabulary, which is built manually in advance. The positional embedding corresponds to the position of the token in the sentence. Segmentation embedding is used to distinguish which sentence the token belongs to, which is effective in the case where there are multiple input sentences.

Upon obtaining the embedding ${\boldsymbol E}$, we use Transformer
encoder layers\cite{Transformer} to serve as the semantic encoder, enhancing the compression of the data for transmission. This semantic encoder incorporates a multi-head attention mechanism, allowing it to merge each input ${\boldsymbol E}$ with contextual information, thereby capturing a more comprehensive range of semantic details. The multi-head attention mechanism empowers the model to evaluate the entire context of the input, fostering a deeper comprehension of the content. Within each attention head, a self-attention process is executed on the embeddings. This operation involves comparing each piece of the input data to every other piece, thus enabling the model to weigh the importance of each part of the input relative to the rest. By scaling and normalizing these attention scores, the model effectively determines which parts of the input are most relevant in a given context. The output of these Transformer encoder layers is a semantically enriched representation, ${\boldsymbol M}$, which can be expressed as
\begin{eqnarray}\label{}
	{\boldsymbol{M}} = {E_{\boldsymbol{\alpha_T}}}({\boldsymbol{X}}|T=\rm{SA}) = Trans({\boldsymbol{E}}),
\end{eqnarray}
where $\rm{Trans}(\cdot)$ is the Transformer encoders, and $T=\rm{SA}$ means the oriented task is sentiment analysis. The encoded semantics ${\boldsymbol M}$ will be further encoded by the channel encoder and transmitted via the wireless channel.

Considering the sentiment analysis task here, we use the classifier as the semantic decoder, which can be realized via a multi-layer perceptron (MLP). Thus, the sentiment analysis results can be obtained via
\begin{eqnarray}\label{}
	\boldsymbol{p}_l = \rm{MLP}({\boldsymbol{M}'}),
\end{eqnarray}
where $\boldsymbol{M}'$ is the received semantics and $\boldsymbol{p}_l$ represents the predicted probability vector of the $l$-th sample.

The cross-entropy loss function is adopted in this work, which can be expressed as
\begin{eqnarray}\label{}
	{\mathcal{L}_{\rm{SA}}}= - \frac{1}{N}\sum\limits_{l = 1}^N {\sum\limits_{m = 1}^M {{q_{lm}}\log \left( {{p_{im}}} \right)} },
\end{eqnarray}
where $N$ is the number of samples, and $M$ is the number of classes. $q_{lm}$ and $p_{lm}$ are the $m$-th element of label vector ${\boldsymbol q_l}$ and predicted vector ${\boldsymbol p_l}$, respectively.

\subsection{Question Answering}
QA is a critical component of many human-computer interaction systems, and is used in various fields such as information retrieval, and knowledge management. However, in a QA-oriented communication system, the transmitter need to transmit a large amount of original text to the receiver to search for the answer. This can result in a significant transmission burden and may affect the task performance in resource-constraint scenarios. To address this issue, we apply the proposed TESC scheme to the QA task to reduce the transmission burden and improve the task performance.


\begin{figure}
\centering
\includegraphics[width=1\linewidth]{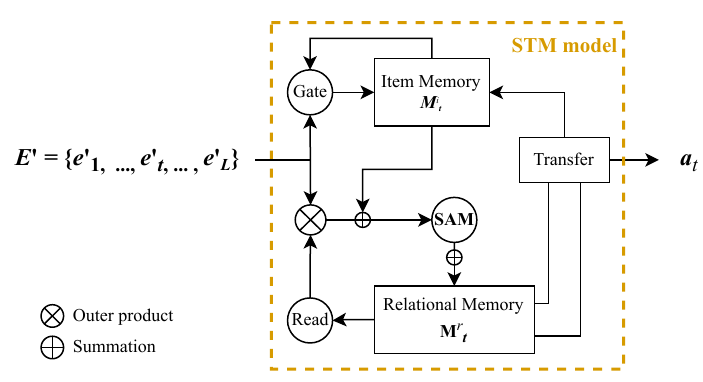}
\caption{The structure of STM model.}
\label{fig4:qamodel}
\end{figure}

The proposed QA-oriented TESC comprises of semantic representation (described in Sec. III), semantic encoder, and semantic decoder. In this subsection, we focus on the semantic codec used specifically for QA-oriented TESC. Given the limited computation and storage resources of the transmitter, it is necessary to simplify the structure of the semantic encoder to strike a balance between the amount of transmitted data and the resource consumption of the transmitter. To this end, we use a simple semantic encoder which primarily consists of an embedding layer that maps the triplets into embedding vectors, which can be optimized during the end-to-end training. This design allows the proposed semantic communication scheme to be easily deployed on various lightweight devices (e.g., IoT devices). The encoded semantics, ${\boldsymbol E}$, can be denoted by
\begin{eqnarray}\label{}
	{\boldsymbol{E}} = {E_{\boldsymbol{\alpha_T}}}({\boldsymbol{X}}|T=\rm{QA}) = \bf{EMB}(\boldsymbol{X}),
\end{eqnarray}
where $\bf{EMB}(\cdot)$ denotes the embedding layer, and $\boldsymbol{X}$ is the transmitted triplets. ${\boldsymbol{E}} = \{{\boldsymbol{e}_1}, {\boldsymbol{e}_2}, ..., {\boldsymbol{e}_l}, ...{\boldsymbol{e}_L}\}$, where $L$ is the number of triplets in $\boldsymbol{X}$ and ${\boldsymbol{e}_l}$ is the encoded semantics corresponding to the $l$-th triplets.

Then, the encoded semantics ${\boldsymbol E}$ will be further encoded by the channel encoder and transmitted via the wireless channel. After obtaining the received signal, the receiver will decode it via channel decoder to recover the semantics ${\boldsymbol E'} = \{{\boldsymbol{e}'_1}, {\boldsymbol{e}'_2}, ..., {\boldsymbol{e}'_l}, ...{\boldsymbol{e}'_L}\}$. 

The semantic decoder utilizes a self-attentive associative memory (SAM)-based two-memory model (STM)\cite{SAM}, which has demonstrated high performance on memory and inference tasks. Inspired by the item and relational memory systems in the human brain, STM separates the relational memory from the item memory. To maintain a detailed representation of the item relationships, the relational memory has a higher-order capacity than item memory, storing multiple relationships represented by matrices instead of scalar or vector quantities. Additionally, the two separate memories interact with each other to enhance their representations. SAM is applied in STM to transform a second-order item memory into a third-order relational representation.

The basic structure of STM model is shown in Fig. \ref{fig4:qamodel}. \mbox{${\mathcal M_t^i}$ $\in$ ${\mathcal {\mathbbm{R}}^{d\times d}}$} is a memory unit for items and ${\mathcal M_t^r}$ $\in$ ${\boldsymbol {\mathbbm{R}}^{n_q\times d\times d}}$ is for relationships. From a high-level view, at $t$-th step, we input ${\boldsymbol e'_t}$ and the previous state of memories \{${\mathcal M_{t-1}^i}$, ${\mathcal M_{t-1}^r}$\} to produce output ${\boldsymbol a_t}$ and new state of memories \{${\mathcal M_t^i}$, ${\mathcal M_t^r}$\}. 

At each step, the item memory ${\mathcal M_t^i}$ is updated with new input ${\boldsymbol e'_t}$ using gating mechanisms. For an input ${\boldsymbol e'_t}$, we update the item memory as
\begin{equation}
\mathcal{M}_t^i=F_t\left(\mathcal{M}_{t-1}^i, \boldsymbol e'_t\right) \odot \mathcal{M}_{t-1}^i+G_t\left(\mathcal{M}_{t-1}^i, \boldsymbol e'_t\right) \odot E_t,
\end{equation}
where $E_t = f_1(\boldsymbol e'_t) \otimes f_2(\boldsymbol e'_t)$, $f_1$ and $f_2$ are feed-forward neural networks that output $d$-dimensional vectors, ${F_t}$ and ${G_t}$ are forget and input gates, respectively.

The associative memories used to store relationships in $\mathcal{M}_t^r$ allow us to reconstruct previously items by reading the relational memory. This involves a two-step contraction process, where $\boldsymbol v_t^r$ is computed as follows
\begin{equation}
\boldsymbol v_t^r=\operatorname{softmax}\left(f_3\left(\boldsymbol e'_t\right)^{\top}\right) \mathcal{M}_{t-1}^r f_2\left(\boldsymbol e'_t\right),
\end{equation}
where ${f_3}$ is a feed-forward neural network that outputs a $n_q$-dimensional vector.

The item memory, together with the read-out from the relational memory, is passed to the SAM module to generate a new relational representation used to update the relational memory ${\mathcal M_t^r}$, as follows
\begin{equation}
\mathcal{M}_t^r=\mathcal{M}_{t-1}^r+\alpha_1 \operatorname{SAM}\left(\mathcal{M}_t^i+\alpha_2 \boldsymbol v_t^r \otimes f_2\left(\boldsymbol e'_t\right)\right),
\end{equation}
where ${\alpha_1}$ and ${\alpha_2}$ are hyper-parameters. The input for SAM is a combination of the current item memory ${\mathcal M_t^i}$ and the association between the extracted item from the previous relational memory ${\boldsymbol v_t^r}$ and the current input data ${\boldsymbol e'_t}$.

The relational memory transfers its knowledge to the item memory by using high dimensional transformation
\begin{equation}
\mathcal{M}_t^i=\mathcal{M}_t^i+\alpha_3 \mathcal{B}_1 \circ \mathcal{V}_f \circ \mathcal{M}_t^r,
\end{equation}
where $\mathcal{V}_f$ is a function that flattens the first two dimensions of its input tensor, $\mathcal{B}_1$ is a feed-forward neural network that maps $\mathbb{R}^{\left(n_q d\right) \times d} \rightarrow \mathbb{R}^{d \times d}$ and ${\alpha_3}$ is a hyper-parameter.

As for the answer to the question, at each time step, we distill the relational memory into an output answer vector $\boldsymbol a_t \in \mathbb{R}^{n_o}$. We alternatively flatten and apply high-dimensional transformations as follows
\begin{equation}
{\boldsymbol a}_t=\mathcal{B}_3 \circ \mathcal{V}_l \circ \mathcal{B}_2 \circ \mathcal{V}_l \circ \mathcal{M}_t^r,
\end{equation}
where $\mathcal{V}_l$ is a function that flattens the last two dimensions of its input tensor. $\mathcal{B}_2$ and $\mathcal{B}_3$ are two feed-forward neural networks that map $\mathbb{R}^{n_q \times(d d)} \rightarrow \mathbb{R}^{n_q \times n_r}$ and $\mathbb{R}^{n_q n_r} \rightarrow \mathbb{R}^{n_o}$ , respectively, where ${n_r}$ is a hyper-parameter.

\section{Experiments}
\label{sec:simulation}

\subsection{Experimental Settings}
\textit{1) Datasets:} The adopted dataset for sentiment analysis is the SST2 dataset\cite{SST2}, which is always used to predict sentiment from long movie reviews. The adopted dataset for QA is the bAbI dataset\cite{bAbI}, which consists of 20 subtasks. For each subtask, there are 1000 questions for training, and 1000 for testing. {Although the SST-2 and bAbI datasets are primarily utilized within the realms of sentiment analysis and QA tasks, respectively, the fundamental NLP capabilities are critical for intelligent IoT systems.}

\begin{table}[t]
	\normalsize
	\vspace{-0.1cm}
	\centering
	\caption{Training Parameters.}
	\setlength{\abovecaptionskip}{-0.5cm}
	\begin{tabular}{ccc}
		\toprule
		Parameter & Sentiment Analysis & QA \\
		\hline
		Epochs & 5 & 300 \\
		Batch size & 8 & 128\\
		Optimizer & Adam & Adam\\
		Learning rate & $1\times 10^{-5}$ & $6\times 10^{-4}$\\
		Drop & 0.3 & 0.5\\
		\toprule
	\end{tabular}
	\vspace{-0.cm}
	\label{tab:experiment}
\end{table}

\begin{table*}[t]
\centering
\caption{{Settings of TESC Network for Different Tasks.}}
\begin{tabular}{|c|ccc|ccc|}
\hline
     Tasks   & \multicolumn{3}{c|}{Sentiment Analysis}                                                                          & \multicolumn{3}{c|}{Question Answeting}                                                                                    \\ \hline
    Module  & \multicolumn{1}{c|}{Layer Name}             & \multicolumn{1}{c|}{Units}              & Activation          & \multicolumn{1}{c|}{Layer Name}                       & \multicolumn{1}{c|}{Units}               & Activation              \\ \hline
\multirow{3}{*}{\begin{tabular}[c]{@{}c@{}}Transmitter\\ (Encoder)\end{tabular}} & \multicolumn{1}{c|}{Embedding Layer}        & \multicolumn{1}{c|}{768}                & Linear                   & \multicolumn{1}{c|}{\multirow{3}{*}{Embedding Layer}} & \multicolumn{1}{c|}{\multirow{3}{*}{64}} & \multirow{3}{*}{Linear} \\ \cline{2-4}
        & \multicolumn{1}{c|}{Transformer Encoders}   & \multicolumn{1}{c|}{256 (12 heads)}     & Gelu                     & \multicolumn{1}{c|}{}                                 & \multicolumn{1}{c|}{}                    &                         \\ \cline{2-4}
        & \multicolumn{1}{c|}{Dense}                  & \multicolumn{1}{c|}{192}                & Relu                     & \multicolumn{1}{c|}{}                                 & \multicolumn{1}{c|}{}                    &                         \\ \hline
Channel                                                                          & \multicolumn{1}{c|}{AWGN/Rayleigh}          & \multicolumn{1}{c|}{None}               & None                     & \multicolumn{1}{c|}{AWGN/Rayleigh}                    & \multicolumn{1}{c|}{None}                & None                    \\ \hline
\multirow{3}{*}{\begin{tabular}[c]{@{}c@{}}Receiver\\ (Decoder)\end{tabular}}    & \multicolumn{1}{c|}{Dense}                  & \multicolumn{1}{c|}{192}                & Relu                     & \multicolumn{1}{c|}{STM Model}                        & \multicolumn{1}{c|}{64}                  & Relu                    \\ \cline{2-7} 
        & \multicolumn{1}{c|}{\multirow{2}{*}{Dense}} & \multicolumn{1}{c|}{\multirow{2}{*}{2}} & \multirow{2}{*}{Softmax} & \multicolumn{1}{c|}{Dense}                            & \multicolumn{1}{c|}{90}                  & Relu                    \\ \cline{5-7} 
        & \multicolumn{1}{c|}{}                       & \multicolumn{1}{c|}{}                   &                          & \multicolumn{1}{c|}{Dense}                            & \multicolumn{1}{c|}{177}                 & Softmax                 \\ \hline
\end{tabular}
\label{tab:network}
\end{table*}

\textit{2) Baselines:} For the baselines, we adopt a task-oriented joint source-channel coding system, a deep learning-based semantic communication system\cite{Xie_Deep}, and the traditional communication system for separate source and channel coding.
\begin{itemize}
    \item[$\bullet$] Error-free Transmission: The full, noiseless texts are delivered to the receiver, which will serve as the upper bound. We label this as ``Error\_free" in the simulation figures.
    \item[$\bullet$] Task-oriented DNN-based joint source-channel coding (JSCC): For sentiment analysis task, the joint codec network consists of Bidirectional Long Short-Term Memory (BiLSTM) layers\cite{BiLSTM}. For the QA task, the joint codec network consists of an End-to-end Memory Network (E2EMN)\cite{E2EMN}. Based on task-oriented JSCC, the receiver can directly execute the task based on the received semantic features without reconstructing the original text. We label this benchmark as ``DeepJSCC'' in the simulation figures.
    \item[$\bullet$] Deep semantic communication based on Transformer: The text is first transmitted via the DeepSC method\cite{Xie_Deep}, and reconstructed at the receiver. Subsequently, the reconstructed text is input into the downstream task network to obtain the desired results. We label this benchmark as ``DeepSC'' in the simulation figures.
    \item[$\bullet$] Traditional methods: To perform the source and channel coding separately, we use Huffman coding for source coding, Reed-Solomon (RS) coding for channel coding, and 16-QAM for modulation, and then execute the task based on the recovered text. We label it as ``Huffman+RS" in the simulation figures. It is worth mentioning that traditional communication systems are not the only ones used here. The source coding can also choose L-Z coding and other coding methods, while the channel coding can choose turbo coding, LDPC coding, and other coding methods.
 \end{itemize}

{It is noteworthy that we have adopted 8-bit quantization for the TESC network, aligning with the approach recommended in \cite{Xie_lite}. This choice enhances the compatibility of the TESC with resource-constraint devices.}
{Top-1 accuracy is used to measure the performance. Experiments are performed by the computer with Ubuntu16.04 + CUDA11.0, and the selected deep learning framework is Pytorch.} {The settings of TESC network structure of different tasks can be found in Table \ref{tab:experiment} and the training parameters are summarized in Table \ref{tab:network}.}

\subsection{Experimental Results}

\begin{figure}[t]
\centering
\includegraphics[width=1\linewidth]{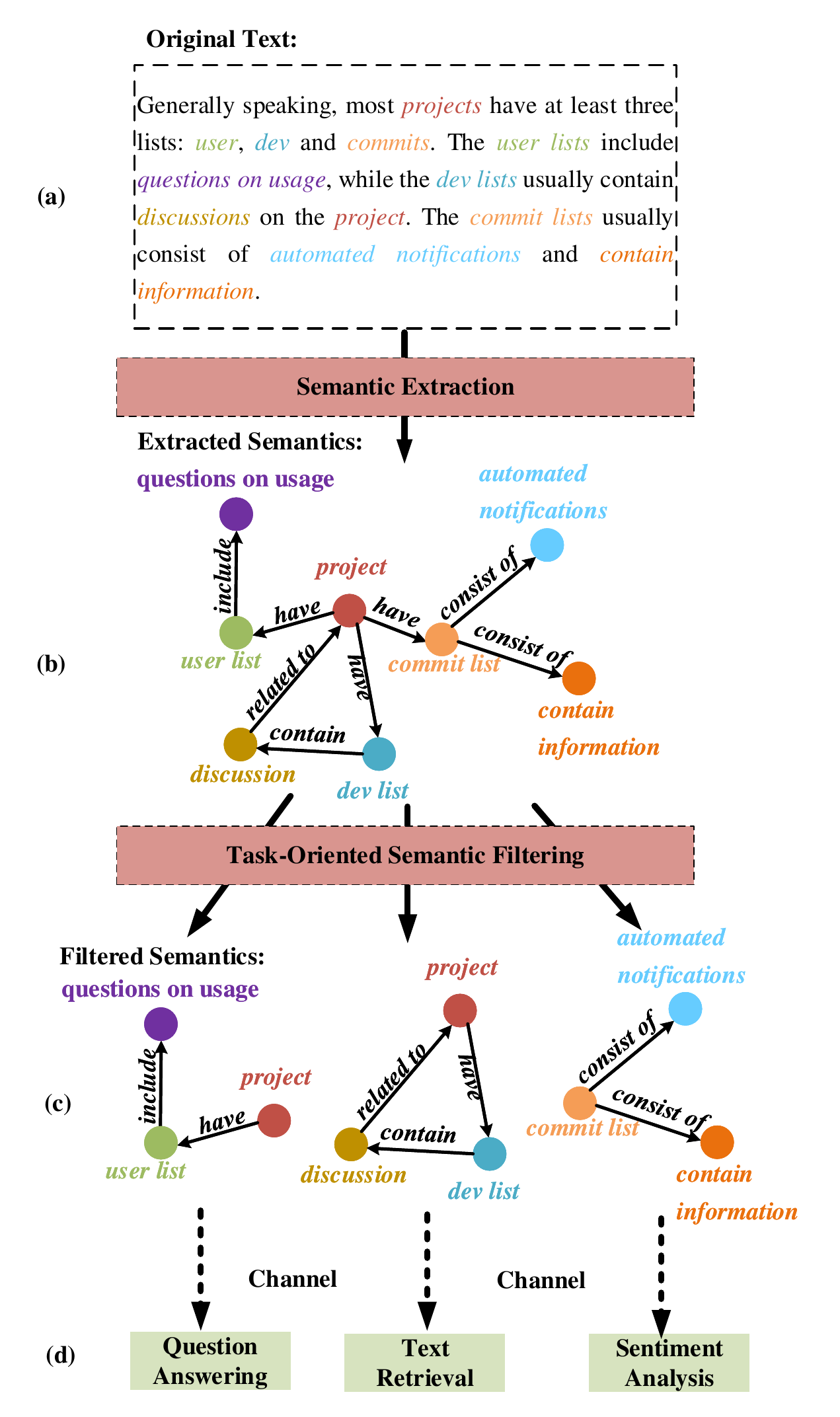}
\caption{An example of the proposed task-oriented TESC scheme.}
\label{example of the system}
\vspace{-0.3cm}
\end{figure}

Fig. \ref{example of the system} shows an example of the proposed task-oriented TESC scheme. In Fig. \ref{example of the system}, the transmitter needs to transmit a text, as shown in Fig. \ref{example of the system}(a), to the receiver to accomplish intelligent tasks. During the communication process, the transmitter first extracts the semantics from the original text, which is represented via some triplets, as shown in Fig. \ref{example of the system}(b). The entities and their corresponding relations are represented as nodes and edges, respectively. Based on the final task, the transmitter then filters the redundant and task-irrelevant triplets to generate transmitted partial task-relevant semantics. Fig. \ref{example of the system}(c) illustrates the filtered results corresponding to various tasks. From Fig. \ref{example of the system}(b) and Fig. \ref{example of the system}(c), we can see that the selected triplets are variable for different tasks, since different tasks focus on different semantic information. This means that the proposed TESC scheme can discard task-irrelevant triplets to reduce the amount of transmitted data. At the receiver, the received semantics can be directly used for implementing the intelligent task, as shown in Fig. \ref{example of the system}(d).

\begin{figure}[t]
\centering
\subfigure[AWGN Channel]{
\includegraphics[width=1\linewidth]{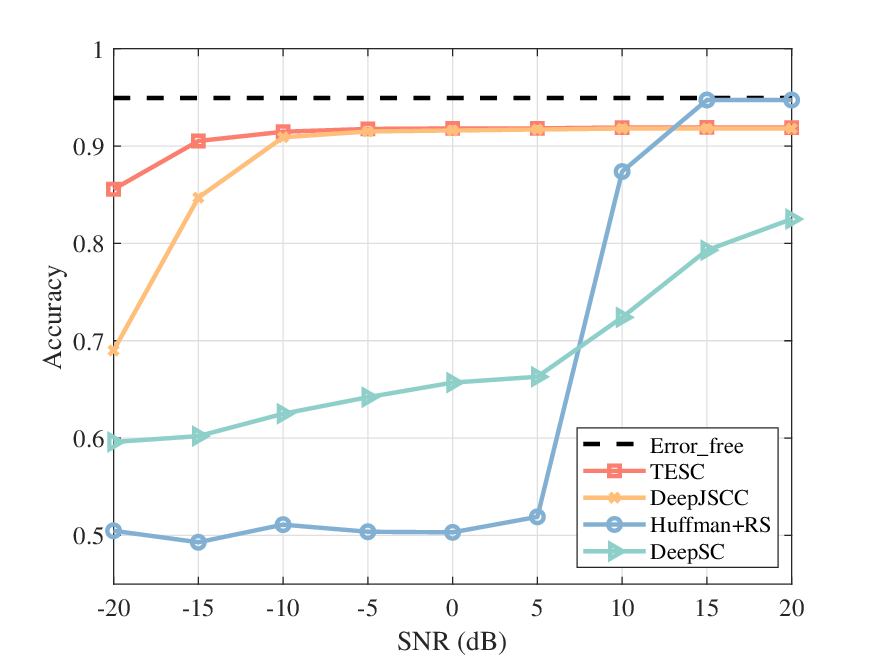}}
\subfigure[Rayleigh Channel]{
\includegraphics[width=1\linewidth]{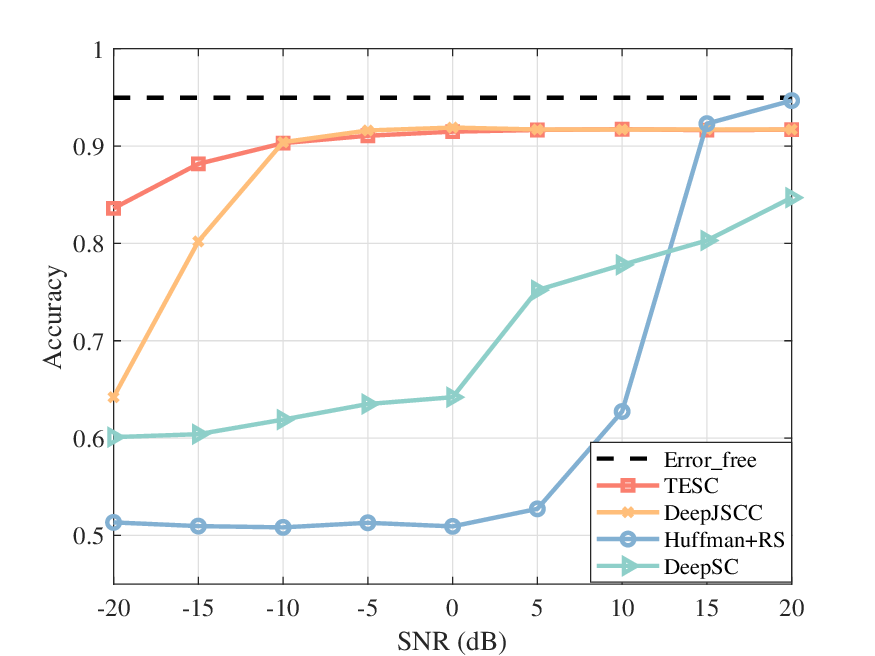}}
\caption{Sentiment analysis accuracy versus SNRs for AWGN and Rayleigh fading channels.}
\label{fig:SA_AWGN}
\vspace{-0.4cm}
\end{figure}
Fig. \ref{fig:SA_AWGN} shows the relationship between sentiment analysis accuracy and signal-to-noise ratio (SNR) in both AWGN and Rayleigh fading channels. As shown in these figures, the accuracy increases with SNR and gradually converges to a certain threshold.  This trend is attributable to higher SNR enhancing communication quality and reducing transmission errors, thus boosting task performance at the receiving end. Notably, the proposed TESC scheme consistently outperforms the DeepJSCC across the entire SNR region. This superior performance of TESC is linked to its ability to capture flawless semantics through triplets, representing text semantics more accurately. Furthermore, TESC's capability to compress semantics tailored to the specific task effectively leverages task-related prior information. It is also evident that both TESC and DeepJSCC surpass DeepSC and traditional communication methods, particularly in lower SNR environments, affirming the efficiency of task-oriented communication strategies. In the context of AWGN channel at 5 dB, TESC scheme demonstrates a significant 80.5\% improvement in accuracy over traditional communication method. This is because deep learning-based methods take the channel condition into consideration in the training process and have better robustness to channel noise. When the SNR exceeds 15 dB, traditional communication methods attain optimal performance. This superior performance is attributed to the fact that, at high SNR levels, traditional communication techniques are capable of achieving error-free transmission, thus ensuring peak accuracy.

\begin{figure}[t]
\centering
\subfigure[AWGN Channel]{
\includegraphics[width=1\linewidth]{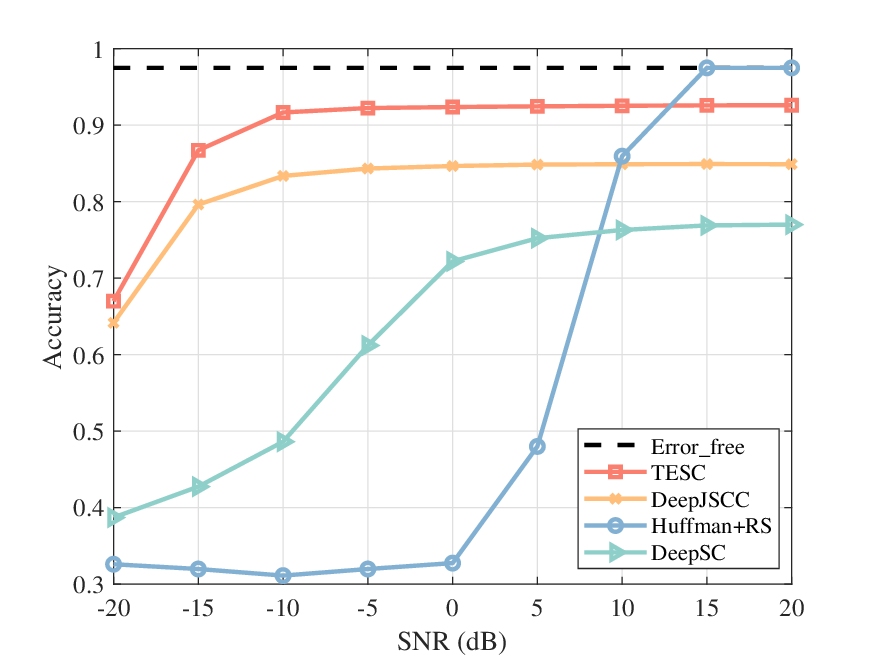}}
\subfigure[Rayleigh Channel]{
\includegraphics[width=1\linewidth]{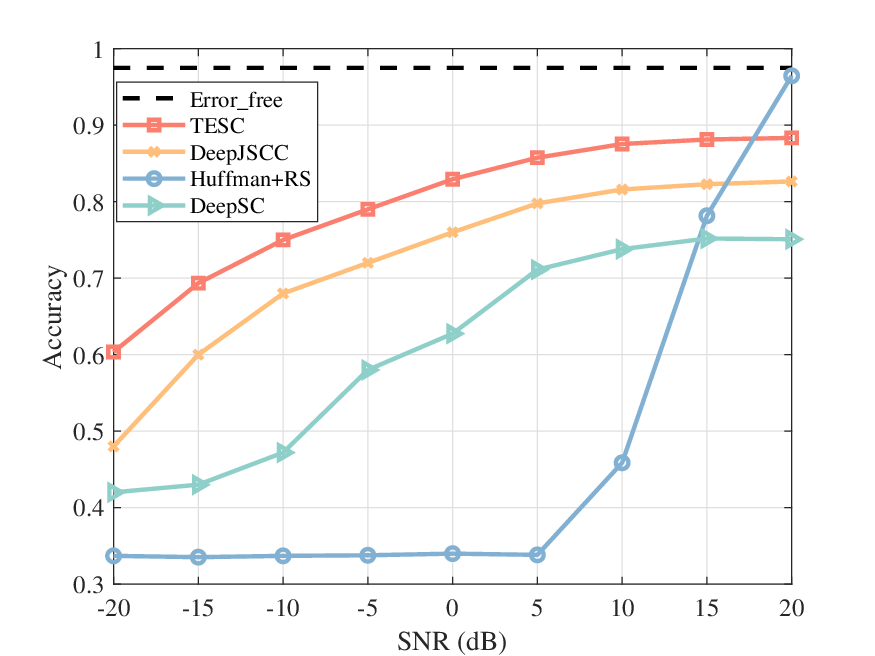}}
\caption{QA accuracy versus SNRs for AWGN and Rayleigh fading channels.}
\label{fig:QA_AWGN}
\vspace{-0.3cm}
\end{figure}
The accuracy of QA task versus SNR over different channels is depicted in Fig. \ref{fig:QA_AWGN}. Echoing the trends observed in Fig. \ref{fig:SA_AWGN}, Fig. \ref{fig:QA_AWGN} demonstrates that accuracy improves with increasing SNR, as higher SNR reduces transmission distortions. In addition, it can be observed that the proposed TESC scheme outperforms DeepJSCC, DeepSC, and traditional communication method, nearing the upper bound at high SNR regime. Specifically, at an SNR of 5 dB in the Rayleigh channel, TESC scheme shows a 7.5\%, 20.6\% and 150\% increase in accuracy compared to DeepJSCC, DeepSC, and traditional method, respectively. This enhanced performance is attributed to TESC's advanced semantic extraction capabilities, which efficiently preserve the most relevant information for the QA task, even in challenging communication environments. Conversely, DeepSC and the traditional method exhibit poorer performance, likely due to the loss of task-relevant semantic information in the recovered text, leading to task failure. Moreover, the traditional method suffers from the \emph{cliff effect}, which results in a sharp decrease in the performance when the channel condition is worse than a threshold. The findings from both Fig. \ref{fig:SA_AWGN} and Fig. \ref{fig:QA_AWGN} lead to the conclusion that TESC scheme not only excels in performance but also possesses a broad applicability across various intelligent tasks.

\begin{figure}
\centering
\includegraphics[width=\linewidth]{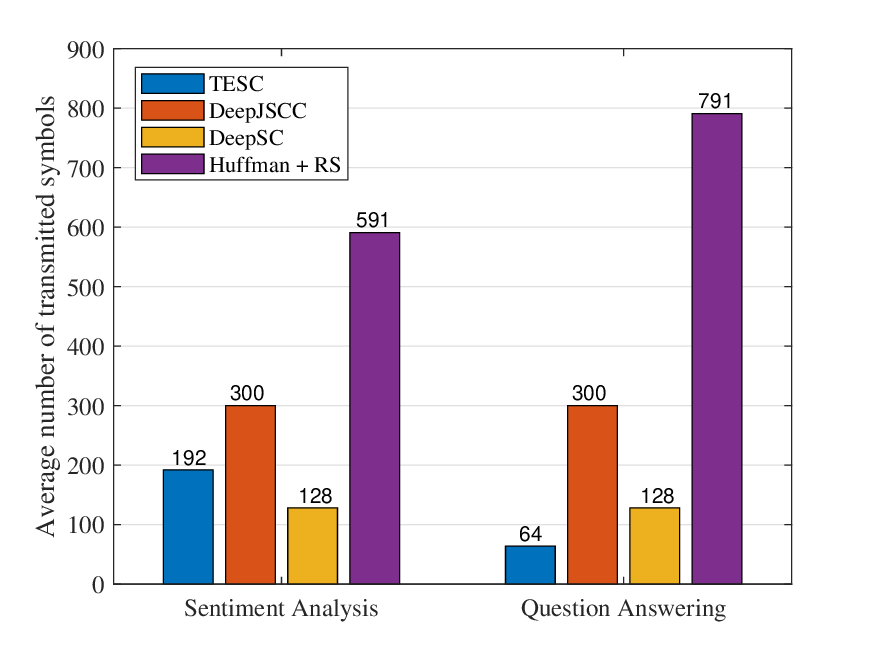}
\caption{The number of transmitted symbols comparison between TESC and various baselines.}
\label{transmitted number of symbols}
\end{figure}

{The numbers of transmission symbols for different methods and various tasks are compared in Fig. \ref{transmitted number of symbols}. All of the displayed results are the average number of symbols transmitted per sentence computed across the entire dataset. It is noteworthy that all of the deep learning-based methods achieve relatively lower numbers of transmission symbols, demonstrating the powerful compression capability of neural networks and highlighting the advantages of semantic communications. Furthermore, the figure clearly shows that the TESC scheme transmits the least symbols compared with DeepJSCC and DeepSC in QA task. This is attributed to the proposed semantic filtering approach, which can efficiently discard redundant and irrelevant semantics, demonstrating the effectiveness of the proposed semantic representation method. In the sentiment analysis task, DeepSC attains the lowest number of transmitted symbols. However, this comes at the cost of increased computational complexity. Consequently, the proposed TESC scheme is particularly suitable for scenarios with limited communication resources.}

\begin{table*}[t]
\normalsize
\centering
\caption{Effectiveness Evaluation of Semantic Filtering Method.}
\begin{tabular}{|c|c|c|c|}
\hline
                          & Average Number of triplets & Average Number of Words & Performance Decay \\ \hline
Before Semantic Filtering & 70.8                                & 517.3                   & -                 \\ \hline
After Semantic Filtering  & 16.9                                & 83.6                    & 0.4\%            \\ \hline
\end{tabular}
\label{tab:filtering_results}
\end{table*}

Table \ref{tab:filtering_results} illustrates the effectiveness of the proposed semantic filtering method. The data illustrates that the semantic filtering method notably decreases the average number of triplets by as much as 76.1\% and the total word count by 83.8\%. Impressively, this reduction only results in a minor 0.4\% dip in task performance, which is essentially negligible. This is attributed to the method's proficiency in eliminating triplets that are either redundant or irrelevant to the task at hand.  Consequently, these results demonstrate that semantic filtering is highly effective in reducing the volume of data transmitted, thereby enhancing communication efficiency without compromising task performance.

\renewcommand\arraystretch{1.5}
\begin{table}[t]
    \normalsize
    \centering
    \caption{{The Comparison Between Proposed TESC and Other Communication Methods in Terms of Flops, Parameters, and Size.}}
    \begin{tabular}{|c|c|c|c|}
    \hline
        Method & FLOPs & Parameters & Size\\ \hline
        TESC & 4.6 $\times 10^7$ & 1,196,078 & 1.14 MB\\ \hline
        DeepJSCC & 4.8 $\times 10^7$ & 1,366,274 & 5.21 MB\\ \hline
        DeepSC & 8.3 $\times 10^7$ & 3,333,120 & 12.3 MB\\ \hline
        Huffman + RS & 9.3 $\times 10^7$ & - & -\\ \toprule
    \end{tabular}
    \label{Computational Complexity}
    \vspace{-0.3cm}
\end{table}

{Table \ref{Computational Complexity} provides a comprehensive comparison between the QA-oriented TESC and baseline communication methods, evaluating computational complexity, parameters, and storage size. As evidenced by the data in this table, all deep learning-based methods demonstrate superior computational efficiency compared to traditional communication methods, with TESC's complexity potentially reduced by more than 50\%. Furthermore, both TESC and DeepJSCC outperform DeepSC and conventional methods in terms of computational complexity. This distinction arises from the former's direct design for intelligent tasks, while the latter require post-text recovery processing, separating transmission and understanding. Notably, the proposed TESC scheme attains the lowest complexity among all the systems under consideration. Additionally, the proposed TESC scheme, requiring only 1.14 MB, achieves substantial space savings through the use of 8-bit parameter quantization, a notable contrast to existing methods relying on 32-bit quantization. In summary, these results collectively affirm the suitability of the proposed TESC scheme for resource-constrained devices.}

\renewcommand\arraystretch{1.2}
\begin{table*}[h]
    \normalsize
    \centering
    \caption{Part of The Results for QA Task at 10 dB Over Rayleigh Channel.}
    \begin{tabular}{|c|c|c|}
    \hline
        \multicolumn{3}{|l|}{\cellcolor[HTML]{F5D9D7}\textbf{Original Text: }\emph{The hallway is east of the bathroom. The bedroom is west of the bathroom.}}\\
        \multicolumn{3}{|l|}{\cellcolor[HTML]{F5D9D7}\textbf{Question: }\emph{What is the bathroom east of?}}\\
        \multicolumn{3}{|l|}{\cellcolor[HTML]{F5D9D7}\textbf{Label: } \emph{bedroom}}\\\hline
        Methods & Recovered Text & Task Result  \\ \hline
        \textbf{TESC} & \textbf{——} & bedroom $\large \color{green}\checkmark$ \\ \hline
        \textbf{DeepJSCC} & \textbf{——} & hallway {\color{red}{\XSolidBrush}} \\ \hline
        \textbf{DeepSC} & the hallway is west of the bathroom. the bathroom is west of the bathroom. & hallway {\color{red}{\XSolidBrush}} \\ \hline
        \textbf{Huffman + RS} & The hallway is east of the bathroom. The bedroom is west of the bathroom. & bedroom $\large \color{green}\checkmark$ \\ \hline
        \multicolumn{3}{|l|}{\cellcolor[HTML]{F5D9D7}\textbf{Original Text: }\emph{The bedroom is west of the kitchen. The hallway is west of the bedroom.}}\\
        \multicolumn{3}{|l|}{\cellcolor[HTML]{F5D9D7}\textbf{Question: }\emph{What is west of the kitchen?}}\\
        \multicolumn{3}{|l|}{\cellcolor[HTML]{F5D9D7}\textbf{Label: } \emph{bedroom}}\\ \hline
        Methods & Recovered Text & Task Result  \\ \hline
        \textbf{TESC} & \textbf{——} & bedroom $\large \color{green}\checkmark$ \\ \hline
        \textbf{DeepJSCC} & \textbf{——} & bedroom $\large \color{green}\checkmark$ \\ \hline
        \textbf{DeepSC} & the bedroom is west of the kitchen. the bedroom is west of the hallway. & bedroom $\large \color{green}\checkmark$ \\ \hline
        \textbf{Huffman + RS} & The bedroohils west of  r  saeechen. The hallway is west of the bedroom. & garden {\color{red}{\XSolidBrush}} \\ \hline
        \multicolumn{3}{|l|}{\cellcolor[HTML]{F5D9D7}\textbf{Original Text: }\emph{The bathroom is north of the garden. The hallway is north of the bathroom.}}\\
        \multicolumn{3}{|l|}{\cellcolor[HTML]{F5D9D7}\textbf{Question: }\emph{What is north of the garden?}}\\
        \multicolumn{3}{|l|}{\cellcolor[HTML]{F5D9D7}\textbf{Label: } \emph{bathroom}}\\\hline
        Methods & Recovered Text & Task Result  \\ \hline
        \textbf{TESC} & \textbf{——} & bathroom $\large \color{green}\checkmark$ \\ \hline
        \textbf{DeepJSCC} & \textbf{——} & bathroom $\large \color{green}\checkmark$ \\ \hline
        \textbf{DeepSC} & the kitchen is south of the bathroom. the kitchen is south of the bedroom. & bedroom {\color{red}{\XSolidBrush}} \\ \hline
        \textbf{Huffman + RS} & The bathroom is north of the garden. Tdehallway is north of the bathrodi. & office {\color{red}{\XSolidBrush}} \\ \hline
    \end{tabular}
    \label{results of QA}
\end{table*}

Some of the visualized results for QA task is displayed in Table \ref{results of QA}. Table \ref{results of QA} presents both the reconstructed text and the results of the QA task for TESC scheme and other baseline methods at a 10 dB Rayleigh channel. The results clearly demonstrate that TESC scheme outperforms others, accurately answering all questions without needing to reconstruct the original text. DeepJSCC ranks second to TESC, highlighting the advantage of focusing on semantic communication for intelligent tasks, as it more effectively preserves essential semantic information pertinent to the task, despite not having access to the original text. While methods like DeepSC and traditional communication approaches can partially recover the original text, they often fail to capture key semantic details critical for the task, as evidenced in their performance in the QA task.

\section{Conclusion}
\label{conclusion}
In this paper, we have proposed the TESC scheme, an explainable task-oriented semantic communication scheme that leverages triplets to represent semantics, enabling the system to handle various text tasks. We have developed a novel semantic representation method that comprises semantic extraction and semantic filtering to transform texts into triplets. The semantic filtering process eliminates redundant or task-irrelevant semantics based on the task-specific knowledge. To evaluate the effectiveness of TESC scheme, we have applied it to sentiment analysis and question answering tasks. Simulation results demonstrate that TESC scheme outperforms various benchmarks, particularly in low SNR environments. Consequently, we believe that TESC scheme is a promising candidate for future task-oriented semantic communication systems. Future extensions of this work include 1) exploring explainable semantic communication for computer vision tasks, and 2) developing unified semantic representation method for multimodal data.

%
\bibliographystyle{IEEEbib}
\nocite{*}\bibliography{stimreference}

\begin{thebibliography}{10}

\bibitem{Liu_Triplet}
C.~Liu, C.~Guo, S.~Wang, Y.~Li, and D.~Hu,
\newblock ``Task-oriented semantic communication based on semantic triplets,''
\newblock in {\em Proc. IEEE Wireless Commun. Netw. Conf. (WCNC)}, Glasgow, United Kingdom, Mar. 2023, pp. 1--6.

\bibitem{Walid_6G}
W.~Saad, M.~Bennis, and M.~Chen,
\newblock ``A vision of 6{G} {W}ireless {S}ystems: Applications, {T}rends,{T}echnologies, and {O}pen {R}esearch {P}roblems,''
\newblock {\em IEEE Netw.}, vol. 34, no. 3, pp. 134--142, June 2020.

\bibitem{Chen_distribute}
M.~Chen, D.~Gündüz, K.~Huang, W.~Saad, M.~Bennis, A.~V. Feljan, and H.~V. Poor,
\newblock ``Distributed {L}earning in {W}ireless {N}etworks: {R}ecent {P}rogress and {F}uture {C}hallenges,''
\newblock {\em IEEE J. Sel. Areas Commun.}, vol. 39, no. 12, pp. 3579 -- 3605, Dec. 2021.

\bibitem{Letaief_Roadmap}
K.~B. Letaief, W.~Chen, Y.~Shi, J.~Zhang, and Y.~A. Zhang,
\newblock ``The {R}oadmap to 6{G}: {AI} {E}mpowered {W}ireless {N}etworks,''
\newblock {\em IEEE Commun. Mag.}, vol. 57, no. 8, pp. 84--90, Aug. 2019.

\bibitem{Qin_survey}
Z.~Qin, X.~Tao, J.~Lu, and G.~Y. Li,
\newblock ``Semantic communications: Principles and challenges,''
\newblock {\em arXiv preprint arXiv:2201.01389}, Jan. 2022.

\bibitem{Nine}
W.~Tong and G.~Y. Li,
\newblock ``Nine challenges in artificial intelligence and wireless communications for 6{G},''
\newblock {\em IEEE Wireless Commun.}, pp. 1--10, 2022.

\bibitem{framework_Yang}
Y.~Yang, C.~Guo, F.~Liu, C.~Liu, L.~Sun, Q.~Sun, and J.~Chen,
\newblock ``Semantic communications with artificial intelligence tasks: reducing bandwidth requirements and improving artificial intelligence task performance,''
\newblock {\em IEEE Ind. Electron. Mag., Early Access}, May 2022.

\bibitem{info}
C.~E. Shannon and W.~Weaver,
\newblock {\em The Mathematical Theory of Communication},
\newblock Champaign, Il, USA: Univ. Illinois Press, 1949.

\bibitem{Xie_Deep}
H.~Xie, Z.~Qin, G.~Y. Li, and B.~Juang,
\newblock ``Deep {L}earning {E}nabled {S}emantic {C}ommunication {S}ystems,''
\newblock {\em IEEE Trans. Signal Process.}, vol. 69, no. 1, pp. 2663--2675, Apr. 2021.

\bibitem{Xie_lite}
H.~Xie and Z.~Qin,
\newblock ``A {L}ite {D}istributed {S}emantic {C}ommunication {S}ystem for {I}nternet of {T}hings,''
\newblock {\em IEEE J. Sel. Areas Commun.}, vol. 39, no. 1, pp. 142--153, Jan. 2021.

\bibitem{Gunduz_JSCC}
E.~Bourtsoulatze, D.~Burth Kurka, and D.~Gunduz,
\newblock ``Deep joint source-channel coding for wireless image transmission,''
\newblock {\em IEEE Trans. Cognit. Commun. Netw.}, vol. 5, no. 3, pp. 567--579, Sep. 2019.

\bibitem{Weng1}
Z.~Weng and Z.~Qin,
\newblock ``Semantic communication systems for speech transmission,''
\newblock {\em IEEE J. Sel. Areas Commun.}, vol. 39, no. 8, pp. 2434--2444, 2021.

\bibitem{retrival}
M.~Jankowski, D.~Gündüz, and K.~Mikolajczyk,
\newblock ``Wireless image retrieval at the edge,''
\newblock {\em IEEE J. Sel. Areas Commun.}, vol. 39, no. 1, pp. 89--100, 2021.

\bibitem{Lee}
C.-H. Lee, J.-W. Lin, P.-H. Chen, and Y.-C. Chang,
\newblock ``Deep learning-constructed joint transmission-recognition for {I}nternet of {T}hings,''
\newblock {\em IEEE Access}, vol. 7, pp. 76547–76561, Jun. 2019.

\bibitem{liu2023adaptable}
C.~Liu, C.~Guo, Y.~Yang, and N.~Jiang,
\newblock ``Adaptable semantic compression and resource allocation for task-oriented communications,''
\newblock {\em IEEE Trans. Cognit. Commun. Netw., Early Access}, 2023.

\bibitem{wei2023federated}
H.~Wei, W.~Ni, W.~Xu, F.~Wang, D.~Niyato, and P.~Zhang,
\newblock ``Federated semantic learning driven by information bottleneck for task-oriented communications,''
\newblock {\em IEEE Commun. Lett.}, 2023.

\bibitem{MU-DeepSC1}
H.~Xie, Z.~Qin, and G.~Y. Li,
\newblock ``Task-oriented multi-user semantic communications for {VQA},''
\newblock {\em IEEE Wireless Commun. Lett.}, Dec. 2021.

\bibitem{Uysal}
E.~Uysal, O.~Kaya, A.~Ephremides, J.~Gross, M.~Codreanu, P.~Popovski, M.~Assaad, G.~Liva, A.~Munari, and B.~Soret,
\newblock ``Semantic communications in networked systems: A data significance perspective,''
\newblock {\em IEEE Netw.}, vol. 36, no. 4, pp. 233--240, 2022.

\bibitem{Kountouris}
M.~Kountouris and N.~Pappas,
\newblock ``Semantics-empowered communication for networked intelligent systems,''
\newblock {\em IEEE Commun. Mag.}, vol. 59, no. 6, pp. 96--102, 2021.

\bibitem{Ma_explainable}
S.~Ma, W.~Qiao, Y.~Wu, H.~Li, G.~Shi, D.~Gao, Y.~Shi, S.~Li, and N.~Al-Dhahir,
\newblock ``Task-oriented explainable semantic communications,''
\newblock {\em IEEE Trans. Wireless Commun., Early Access}, Apr. 2023.

\bibitem{Jiang_KG}
S.~Jiang, Y.~Liu, Y.~Zhang, P.~Luo, K.~Cao, J.~Xiong, H.~Zhao, and J.~Wei,
\newblock ``Reliable semantic communication system enabled by knowledge graph,''
\newblock {\em Entropy}, vol. 22, no. 6, pp. 846, 2022.

\bibitem{Hu_robust}
L.~Hu, Y.~Li, H.~Zhang, L.~Yuan, F.~Zhou, and Q.~Wu,
\newblock ``Robust semantic communication driven by knowledge graph,''
\newblock in {\em 9th International Conference on Internet of Things: Systems, Management and Security (IOTSMS)}, Milan, Italy, Nov. 2022, pp. 1--5.

\bibitem{Wang_Attention2}
Y.~Wang, M.~Chen, T.~Luo, W.~Saad, D.~Niyato, H.~V. Poor, and S.~Cui,
\newblock ``Performance {O}ptimization for {S}emantic {C}ommunications: {A}n {A}ttention-based {R}einforcement {L}earning {A}pproach,''
\newblock {\em IEEE J. Sel. Areas Commun.}, vol. 40, no. 9, pp. 2598 -- 2613, Sept. 2022.

\bibitem{niu2022paradigm}
K.~Niu, J.~Dai, S.~Yao, S.~Wang, Z.~Si, X.~Qin, and P.~Zhang,
\newblock ``A paradigm shift toward semantic communications,''
\newblock {\em IEEE Commun. Mag.}, vol. 60, no. 11, pp. 113--119, 2022.

\bibitem{openie1}
M.~Banko, M.~J. Cafarella, S.~Soderland, M.~Broadhead, and O.~Etzioni,
\newblock ``Open information extraction from the web,''
\newblock {\em Communications of the ACM}, vol. 51, no. 12, pp. 68--74, 2008.

\bibitem{Shi}
G.~Shi, Y.~Xiao, Y.~Li, and X.~Xie,
\newblock ``From semantic communication to semantic-aware networking: {M}odel, architecture, and open problems,''
\newblock {\em IEEE Commun. Mag.}, vol. 59, no. 8, pp. 44--50, 2021.

\bibitem{openie2}
G.~Angeli, M.~J.~J. Premkumar, and C.~D. Manning,
\newblock ``Leveraging linguistic structure for open domain information extraction,''
\newblock in {\em Proceedings of the 53rd Annual Meeting of the Association for Computational Linguistics and the 7th International Joint Conference on Natural Language Processing}, 2015, pp. 344--354.

\bibitem{Che_KG}
Y.~Che, H.~Xiong, S.~Han, and X.~Xu,
\newblock ``Cache-enabled {K}nowledge {B}ase {C}onstruction {S}trategy in {S}emantic {C}ommunications,''
\newblock in {\em Proc. IEEE Global Commun. Conf. (GLOBECOM)}, Rio de Janeiro, Brazil, Dec. 2022.

\bibitem{sentiment}
W.~Medhat, A.~Hassan, and H.~Korashy,
\newblock ``Sentiment analysis algorithms and applications: {A} survey,''
\newblock {\em Ain Shams engineering journal}, vol. 5, no. 4, pp. 1093--1113, 2014.

\bibitem{QA}
L.~Hirschman and R.~Gaizauskas,
\newblock ``Natural language question answering: the view from here,''
\newblock {\em natural language engineering}, vol. 7, no. 4, pp. 275--300, 2001.

\bibitem{nlp}
E.~Cambria and B.~White,
\newblock ``Jumping nlp curves: A review of natural language processing research,''
\newblock {\em IEEE Computational intelligence magazine}, vol. 9, no. 2, pp. 48--57, 2014.

\bibitem{Transformer}
A.~Vaswani, N.~Shazeer, N.~Parmar, J.~Uszkoreit, L.~Jones, A.~N. Gomez, Ł. Kaiser, and I.~Polosukhin,
\newblock ``Attention is all you need,''
\newblock in {\em Advances Neural Info. Process. Systems (NIPS’17)}, Long Beach, CA, USA, Dec. 2017, pp. 5998--6008.

\bibitem{SAM}
H.~Le, T.~Tran, and S.~Venkatesh,
\newblock ``{S}elf-{A}ttentive {A}ssociative {M}emory,''
\newblock in {\em International Conference on Machine Learning (ICML)}, online, Jul. 2020.

\bibitem{SST2}
R.~Socher, A.~Perelygin, J.~Wu, J.~Chuang, C.~D. Manning, A.~Ng, and C.~Potts,
\newblock ``Recursive deep models for semantic compositionality over a sentiment treebank,''
\newblock in {\em Proceedings of the 2013 conference on empirical methods in natural language processing}, 2013, pp. 1631--1642.

\bibitem{bAbI}
J.~Weston, A.~Bordes, S.~Chopra, A.~M. Rush, B.~van Merriënboer, A.~Joulin, and T.~Mikolov,
\newblock ``Towards {A}i-{C}omplete {Q}uestion {A}nswering: {A} {S}et of {P}rerequisite {T}oy {T}asks,''
\newblock {\em arXiv preprint arXiv:1502.05698}, Mar. 2015.

\bibitem{BiLSTM}
S.~Zhang, D.~Zheng, X.~Hu, and M.~Yang,
\newblock ``Bidirectional long short-term memory networks for relation classification,''
\newblock in {\em Proceedings of the 29th Pacific Asia conference on language, information and computation}, 2015, pp. 73--78.

\bibitem{E2EMN}
S.~Sukhbaatar, A.~Szlam, J.~Weston, and R.~Fergus,
\newblock ``End-to-end memory networks,''
\newblock {\em Advances in neural information processing systems}, vol. 28, 2015.

\end{thebibliography}

\end{document}